\documentclass[aps,prb,twocolumn,superscriptaddress,longbibliography]{revtex4-2}
\usepackage[colorlinks=true,linkcolor=blue,citecolor=blue]{hyperref}
\usepackage{amsfonts}
\usepackage{subfigure}
\usepackage{amsmath}
\usepackage{txfonts}
\usepackage{amssymb}
\usepackage{amsbsy}
\usepackage{epsfig}
\usepackage{graphicx}
\usepackage{epstopdf}
\usepackage{mathdots}
\usepackage{color}
\usepackage{cleveref}
\usepackage{bbm}
\usepackage{natbib}
\usepackage{url}
\usepackage[utf8]{inputenc}

\newcommand{\be}{\begin{equation}}
\newcommand{\ee}{\end{equation}}

\newcommand{\bs}{\boldsymbol}

\begin{document}

\title{Laser induced Flqouet topological phases in a 2D weak topological insulator with $\bs{\it unconventional}$ nonlinear dispersion }

\author{Aayushi Agrawal}
\email{p20170415@pilani.bits-pilani.ac.in}
\affiliation{Department of Physics, Birla Institute of Technology and Science, Pilani 333031, India}
\author{Jayendra N. Bandyopadhyay}
\email{jnbandyo@gmail.com}
\affiliation{Department of Physics, Birla Institute of Technology and Science, Pilani 333031, India}

\date{\today}

\begin{abstract}

Recently, we presented a two-dimensional (2D) model of a weak topological insulator formed by stacking an $N$ number of Su-Schrieffer-Heeger (SSH) chains \cite{Agrawal_2022-02}. We now study the influence of periodic driving on the topological properties of this system, which has all the fundamental symmetries, by shining it with circularly polarized light (CPL). The CPL is chosen because it breaks the time-reversal symmetry, which induces more exotic topological properties in the system. We investigate two different formations of the $N$ stacked SSH chains: all the SSH chains are topologically trivial in one formation and nontrivial in the other one. In contrast to the undriven or static case, both formations exhibit distinct topological behaviors. Here, we particularly derive the Floquet or the effective Hamiltonian using the replica method, which facilitates the study of high- and low-frequency regimes. We have discovered that this model exhibits laser-induced Floquet topological phases with higher Chern numbers. This system has nonlinear dispersion along both directions with additional $k_x-k_y$ coupling terms, which made the dispersion of this system {\it unconventional}. We closely study the role of this unconventional dispersion in the system at the low-energy limit and its response to periodic driving. The low-energy Hamiltonian also reveals a hierarchy in the gaps of the neighboring Floquet bands. Interestingly, though this model has nonlinear quasi-energy dispersion, it still shows some signatures of hierarchy, which was observed in the system with linear dispersion like graphene. Furthermore, we study the effect of linearly polarized light (LPL) on the topological properties of the system. In response to the LPL driving, the band-touching point either opens up or splits into two band-touching points.

\end{abstract}

\maketitle

\section{\label{Sec:Introduction}Introduction}

Floquet engineering is becoming a vast area of research, where a desired solid-state system is synthesized by periodic driving. A primary goal of this research area is to design a periodic driving protocol to quickly introduce or enhance desired exotic properties in a given system in a very controlled way \cite{Floquet-01}. In the realm of topological insulators (TIs), periodically driven TIs or Floquet TIs have been studied extensively. Periodic driving introduces many exciting features in the TIs, which may not be possible to realize by any static means \cite{Floquet-top-01,Floquet-top-02,Floquet-top-03,Floquet-top-04,Floquet-top-05,Floquet-top-06,Rechtsman2013,Vega2020}. These periodically driven systems can be realized experimentally by ultracold atoms in optical lattices \cite{Jotzu2014,Bloch2005}.

The quasi-nD systems, engineered by stacking numerous nD systems, came into light after the discovery of weak topological insulators (WTIs). These topological materials were discovered in quasi-2D quantum spin Hall systems (QSH) in which 2D QSH layers were stacked and formed a 3D layered structure \cite{KaneMele3D2007,Moore3D2007}. A few studies concentrated on quasi-2D systems, such as the 3D layered structure of graphene and topological crystalline insulators (TCI) \cite{Das2019,Li_2018,Marwa2023,Kim2015,Qian2020,Zou2016,Ion2019,Liu2016,Burkov2011}. The recent literature focuses primarily on 2D and some quasi-2D materials. Very little attention is paid to the quasi-1D systems formed by a systematic stacking of identical 1D systems. Recently, present authors rigorously study an $N$ number of stacked SSH chains \cite{Agrawal_2022-02}. Similarly, some studies have investigated the topological properties of another quasi-1D model, an extended Su-Schrieffer-Heeger model (E-SSH) \cite{li2014topological, Agrawal_2022-01}. The E-SSH model is a 1D SSH chain where all the hopping amplitudes are modulated by a cyclic parameter $\theta$, and this cyclic parameter is considered as another synthetic dimension. The Floquet version of the SSH chain, the simplest 1D TI, displays interesting properties \cite{FSSH-01, FSSH-02, FSSH-03,FSSH-04}. An intriguing feature of the static E-SSH model is that the system's phase diagram resembles the Haldane model's phase diagram. The interlink between the E-SSH model and the $N$ stacked SSH model is that by promoting $\theta$ to an actual dimension, one can obtain a $N$ stacked SSH model. The E-SSH model is also studied with different periodic driving schemes such as Dirac-delta kickings, and sinusoidal driving \cite{Agrawal_2022-01,FSSH-03}.

Through Floquet, the quasi-2D system of $N$ stacked graphene layers is also explored where circularly polarized light (CPL) is used as an external periodic drive and topological phases with the high Chern number ($C$) are observed \cite{Li_2018}. The CPL is widely recognized for breaking the time-reversal symmetry \cite{CPL-01,CPL-02,CPL-03,CPL-04,CPL-05,CPL-06,CPL-07,FoaTorres2015}. As a result, it creates a new gap at the band touching points, and the system harbors the quantum Hall effect without using any external magnetic field or creating Landau levels \cite{Haldane}. Furthermore, the CPL can be used to detect optical chirality, thermoelectric transport, photo-voltage, the dynamical Hall effect, and as a probe for high harmonic generation \cite{Huang2021,Karch2010,Hatano2009,Tahir2015,Baykusheva2021}.
 
This work aims to induce new Floquet topological phases through CPL in the $N$ stacked SSH chains model. As per our knowledge, the periodically driven 2D WTI is not studied in the literature. Here, we study a 2D WTI under periodic driving. This undriven $N$ stacked SSH chains follows all three fundamental symmetries (chiral, particle-hole, and time-reversal), which host nontrivial topology but with Chern number $C=0$. In the undriven case, two possible constructions were addressed: $N$ stacked SSH model where each SSH chain is topologically trivial (winding number, $w=0$), and each SSH chain is nontrivial ($w=1$) \cite{Agrawal_2022-02}. These two cases exhibit different topological phases as we turn on the driving. This model has an interesting feature because there exists a coupling term between the momenta $k_x$ and $k_y$, which makes the system complex and distinct from various other well-studied 2D systems \cite{BHZ_model2006,QWZ_model2006}. At the low energy limit, near the band touching points, the dispersion relation of this system becomes quadratic with a $k_x k_y$ like coupling term. 
 
In various studies, it has been observed that, in the case of linear dispersion, the band crossing points or Dirac points are stable against any small perturbation \cite{Hatsugai2006,Haldane,Yu2011,Castro2009}. However, the quadratic dispersion near the band crossing points or semi-Dirac points is unstable for the small perturbation: either a gap is opened or splits into two Dirac points \cite{Sun2009,CPL-06}. The emergence of the semi-Dirac point requires linear dispersion in one direction and quadratic dispersion in the other. However, the dispersion relation of the $N$ stacked SSH chains model is {\it unconventional} with nonlinear dispersion along both directions with additional $k_x-k_y$ coupling terms. This unconventional dispersion relation encourages us to investigate the response of the $N$ stacked SSH chains model under the influence of linearly polarized light (LPL) along the $x-$ and $y-$ directions. 

This paper is organized as follows: In Sec. \ref{Sec: Model}, we briefly discuss the static Hamiltonian. In the next section, Sec. \ref{Sec: Driven Hamiltonian}, we discuss the Floquet formalism and the periodically driven Hamiltonian, which uses the exact Floquet replica method. In the next section, Sec. \ref{Sec:Low-energy-H}, we calculate the low-energy Hamiltonian to show a signature of hierarchy in Floquet band gaps. In Sec. \ref{Sec:LPL}, we demonstrate the role of linearly polarized light on the $N$ stacked SSH model. Finally, we summarize in Sec. \ref{Sec:Discussion}.

\section{\label{Sec: Model}Static Hamiltonian}

We study the effect of periodic driving on a $N$ stacked SSH chain model, which is composed of an $N$ number of stacked SSH chains  \cite{Agrawal_2022-02}. In real space, the mathematical expression for this static system is given as
\begin{equation}
\begin{split}
H_{N-SSH}&= (1-\eta)\sum_{n_x,n_y} c_{n_x, n_y}^{\dagger A} c_{n_x, n_y}^B + (1+\eta)\sum_{n_x,n_y} c_{n_x+1, n_y}^{\dagger A} c_{n_x, n_y}^B \\
& +\frac{\delta}{2}\, \sum_{n_x,n_y} \bigg[c_{n_x, n_y}^{\dagger A} c_{n_x, n_y+1}^B + c_{n_x, n_y+1}^{\dagger A} c_{n_x, n_y}^B \bigg]\\
& -\frac{\delta}{2}\, \sum_{n_x,n_y} \bigg[c_{n_x+1, n_y}^{\dagger A} c_{n_x, n_y+1}^B + c_{n_x+1, n_y+1}^{\dagger A} c_{n_x, n_y}^B \bigg] + h.c.
\end{split}
\end{equation}
Here, $\delta$ is the hopping amplitude between the inter-sublattices of neighboring SSH chains, and $\eta$ is the dimerization constant of the individual SSH chain. The parameter $\eta$ can be negative or positive values, which results in a trivial or nontrivial SSH chain. To illustrate the energy spectrum and topological properties, we write the Hamiltonian in the quasimomentum space (or $\bs{k}$-space), which is given as
\begin{equation}
H_{N-SSH}(\bs{k}) = \bs{h} \bs{\cdot} \bs{\sigma}
\label{Eq-NSSH}
\end{equation}
\begin{align*}
h_x(\textbf{k}) =& [(1+\cos k_x) + (1-\cos k_x) \, (\delta \cos k_y - \eta)] \\
h_y(\textbf{k}) =& [(1+\eta) - \delta \cos k_y] \, \sin k_x 
\end{align*}
In our previous study \cite{Agrawal_2022-02}, we found that breaking chiral and time-reversal symmetries is essential for this static system to be a Chern insulator (CI). However, this static system follows all three fundamental symmetries and shows topological properties even when the Chern number $C=0$. For this case, its topological property is determined by nonzero 2D Zak phase $Z(k_y) = -\pi$. This study also revealed that the topological properties of this system are not dependent on the topological properties of the individual SSH chain. We now discuss this model under the periodic driving with CPL. It is well-known that the CPL breaks the chiral and the TR symmetry in the system, leading to nontrivial topological properties in the system with nonzero Chern numbers.

\section{\label{Sec: Driven Hamiltonian}Driven Hamiltonian} 

We now apply a laser field whose vector potential $\bs{A}(t)$ has the form 
\[\bs{A}(t) = (A_{0x} \cos \Omega t, A_{0y} \sin \Omega t)\]
with it satisfies $\bs{A}(t+T) = \bs{A}(t)$, where $T$ is the time-period of the driving, and consequently the driving frequency $\Omega = 2\pi/T$. Here, $A_{0x}$ and $A_{0y}$ are the components of the vector potential along $x$ and $y$ direction, respectively. If we set $A_{0x} = A_{0y} = A_0$, the laser field will be the CPL. The LPL is a special case of this laser field, when it has a form either $\bs{A}(t) = (A_{0x} \cos \Omega t, 0)$ or $\bs{A}(t) = (0, A_{0y} \cos \Omega t)$.
This driving is induced in the system by the Peierls substitution, which modifies the form of the quasi-momenta $k_x$ and $k_y$ as 
\[k_x(t) \rightarrow k_x + A_x(t)\, ; \, k_y(t) \rightarrow k_y + A_y(t) \]\\
The form of the time-periodic Hamiltonian in the $\bs{k}$-space reads
\begin{equation}
\begin{split}
H_{N-SSH}(\textbf{k},t) =& [(1+\cos k_x(t)) + (1-\cos k_x(t)) \\
& \times (\delta \cos k_y(t) - \eta)] \sigma_x \\
&+ [(1+\eta) - \delta \cos k_y(t)] \, \sin k_x(t) \, \sigma_y
\end{split}
\label{Eq-Driven-NSSH}
\end{equation}
In order to solve the time-periodic equation, we use the Floquet replica method and calculate the effective Hamiltonian, an infinite dimensional matrix in the frequency space.

\subsection{Floquet theory}

Periodically driven systems are studied under the Floquet formalism. Therefore, these systems are also known as Floquet systems \cite{Floquettheory1883,shirley1965solution}. Using this theorem, we solve the time-periodic Schr\"odinger equation \cite{shirley1965solution,Eckardt_2015}
\begin{equation}
 i \hbar \, \frac{d}{dt} |\psi(t)\rangle = H(t) \, |\psi(t)\rangle,
\end{equation}
where $H(t) = H(t+T)$. The Floquet theorem is the temporal version of the well-known Bloch's theorem of solid-state physics. Therefore, according to this theorem, the solution of the time-periodic Schr\"odinger equation can be written as
\begin{equation}
|\psi_n(t)\rangle = e^{-i \epsilon_n t} \, |u_n(t)\rangle,
\end{equation}
where $n$ presents the Floquet band index, and $\epsilon_n$ is corresponding quasienergy. The states $|u_n(t)\rangle$ are called Floquet modes, which are periodic in time with the same period as the Hamiltonian, i.e., $|u_n(t+T)\rangle = |u_n(t)\rangle$. The Floquet states are the eigenstates of the single period time-evolution operator, and therefore
\begin{equation}
\hat{U}(t_0+T,t_0) |\psi_n(t_0) \rangle = e^{-i \epsilon_n T} |\psi_n(t_0)\rangle.
\end{equation}
Solving the above eigenvalue problem, one can obtain the Floquet states and the corresponding quasienergies. However, there is an alternative way to calculate the Floquet states and quasienergies by substituting these solutions in the time-periodic Schr\"odinger equation, which eventually takes the following form
\begin{equation}
\left[H(t) - i \frac{\partial}{\partial t}\right] \, |u_n(t)\rangle = \epsilon_n \, |u_n(t)\rangle.
\label{eq:TPSE}
\end{equation}
Due to the time periodicity in $H(t)$ and $u_n(t)$, we can expand these in Fourier series as
\[H(t) = \sum_m \, e^{-im\Omega t} \, H^{(m)} \,;~~ \, |u_n(t)\rangle = \sum_m \, e^{-im\Omega t} \, |u_n^{(m)}\rangle, \]
where $m = 0, \pm 1, \pm 2, \dots$. The bands corresponding to $m=0$ are the central Floquet bands, whereas those with nonzero $m$ form side bands. The central Floquet bands lie in the quasienergy range $-\frac{\Omega}{2} \leq \epsilon \leq \frac{\Omega}{2}$, and this is known as the `first Floquet-Brillouin zone' (FBZ).
The Fourier component $H^{(m)}$ is obtained as
\begin{equation}
H^{(m)} = \frac{1}{T} \int_{0}^{T} H(\bs{k},t) e^{-i m \Omega t} dt.
\end{equation}
In the Fourier space, the time-periodic Schr\"odinger equation given in Eq. \eqref{eq:TPSE} can be written as
\begin{equation}
\epsilon_n |u_n^{(m)}\rangle = \sum_m \left[ H^{m-m^\prime} - m\Omega \delta_{mm^\prime} \right] |u_n^{(m^\prime)}\rangle
\end{equation} 
The above eigenvalue equation corresponds to an infinite dimensional effective Hamiltonian, which is defined in the extended Hilbert space $\mathcal{H} \otimes \mathcal{T}$, where $\mathcal{H}$ is the standard Hilbert space and $\mathcal{T}$ is the Hilbert space which spans all the time-periodic functions $\bigl\{e^{-i m \Omega t}\bigr\}$ \cite{Sambe}. This infinite-dimensional matrix is formed by an infinite number of duplicate copies of undriven energy bands affected by external driving. Therefore, this is known as the ``Floquet replica method", where each copy corresponds to a photon sector.
We obtain quasienergies and the corresponding Floquet modes by diagonalizing the effective Hamiltonian. In numerical calculation, we have to truncate this infinite dimensional matrix and consider only a finite number of photon sectors, where the strength of the driving frequency $\Omega$ decides the number of photon sectors. In the case of the higher frequencies, we need to consider a number of photon sectors for the numerical convergence. On the other hand, for lower frequencies, many photon sectors are to be included in the computation to achieve the desired convergence. The solution of this quasienergy problem is analogous to the dressed atom picture of the laser-atom interaction. Hence, the matrix element or the Fourier component $H^{(m)}$ demonstrates the $m$-photon process \cite{eckardt2008dressed}. 

\subsection{Periodically driven ${\bs N}$ stacked SSH model: Floquet formalism}

First, we write the Fourier components $H^{(m)}$ for the $N$-stacked SSH model in the form as
\[H^{(m)} = \bs{d}^{(m)} \bs{\cdot} \bs{\sigma}. \]
The driving modifies the undriven part as follows
\begin{subequations}
\begin{equation}
H^{(0)}_{N-SSH} = \, \bs{d}^{(0)}_{N-SSH} \cdot \bs{\sigma}
\end{equation}
where
\begin{equation}
\begin{split}
\left(d^{(0)}_{N-SSH}\right)_x &= \, (1-\eta) + (1+\eta) \, \cos k_x \, J_0(A_0) + \delta \, \cos k_y \, J_0(A_0) \\
&- \delta \, \cos k_x \, \cos k_y \, J_0^2(A_0)
\end{split}
\end{equation}
\begin{equation}
 \left(d^{(0)}_{N-SSH}\right)_y = \, (1+\eta) \, \sin k_x \, J_0(A_0) - \delta \, \sin k_x \, \cos k_y \, J_0^2(A_0);
\end{equation}
\end{subequations}
where $J_0$ is the Bessel function of the first kind with zeroth order.

In order to calculate the other non-zero Fourier components, we choose the driving amplitude $A_0$ such that the Bessel functions contribute only up to an order of $A_0^2$. The higher order Fourier components are neglected because we assume that the amplitude $A_0$ is small. Thus, the effective Hamiltonian $H_{\rm eff}$ have only $H^{(1)}$ and $H^{(2)}$ Fourier components, which are given as
\begin{subequations}
\begin{equation}
H^{(1)}_{N-SSH} = \, \bs{d}^{(1)}_{N-SSH} \cdot \bs{\sigma}
\end{equation}
where
\begin{equation}
\begin{split}
\left(d^{(1)}_{N-SSH}\right)_x =& \, -(1+\eta) \, \sin k_x \, J_1(A_0) + i \, \delta \sin k_y \, J_1 (A_0) \\
& -\delta \left\{-\sin k_x \, \cos k_y + i \, \cos k_x \, \sin k_y \right\} J_0(A_0) \, J_1(A_0)
\end{split}
\end{equation}
\begin{equation}
\begin{split}
\left(d^{(1)}_{N-SSH}\right)_y =& \, (1+\eta) \, \cos k_x \, J_1(A_0) \\
& -\delta \left\{\cos k_x \, \cos k_y + i \, \sin k_x \, \sin k_y \right\} J_0(A_0) \, J_1(A_0);
\end{split}
\end{equation}
\label{Eq-F-comp-1}
\end{subequations}
and
\begin{subequations}
\begin{equation}
H^{(2)}_{N-SSH} = \, \bs{d}^{(2)}_{N-SSH} \cdot \bs{\sigma}
\end{equation}
where
\begin{equation}
\begin{split}
\left(d^{(2)}_{N-SSH}\right)_x = & \, -(1+\eta) \, \cos k_x \, J_2(A_0) + \delta \, \cos k_y \, J_2(A_0) \\
& + i \, \delta \sin k_x \, \sin k_y J_1^2(A_0)
\end{split}
\end{equation}
\begin{equation}
 \left(d^{(2)}_{N-SSH}\right)_y = \, - (1+\eta) \, \sin k_x \, J_2(A_0) - i \, \delta \, \cos k_x \, \sin k_y J_1^2(A_0).
\end{equation}
\label{Eq-F-comp-2}
\end{subequations}
Here, $J_1$ and $J_2$ are the Bessel functions of the first kind.
\begin{figure}
\includegraphics[width = 0.45\textwidth]{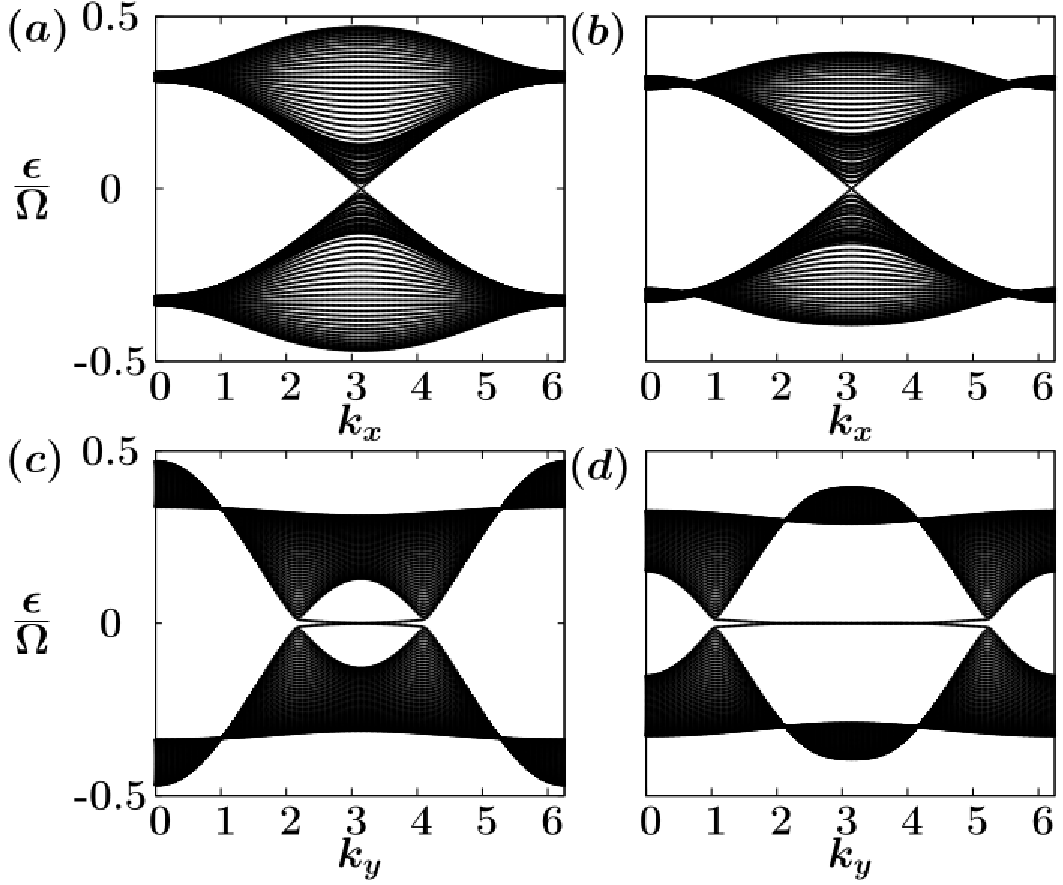}
\caption{Quasienergy bands for the periodically driven $N$ stacked SSH chains are shown in cylindrical geometry. In subfigures (a) and (c), all the SSH chains are considered trivial ($\eta = -0.5$), while in subfigures (b) and (d), all the SSH chains are nontrivial ($\eta = 0.5$). Here, when the quasienergy bands are shown as a function of $k_x$ ($k_y$), then this suggests that the PBC is considered along $x$-direction ($y$-direction), and the OBC is considered along the other direction. In both cases, the driving generates new Floquet topological phases with the Chern number $C = -1$, and the appearance of the edge states is its consequence. We set the parameter values for both plots as $\delta = 1.0$, $A_0 = 0.5$, and $\Omega = 6.0$.}
\label{Fig-01}
\end{figure}

Floquet energy bands corresponding to the Hamiltonian given in Eq. (\ref{Eq-Driven-NSSH}) are shown in Fig \ref{Fig-01}. In subfigure \ref{Fig-01}(a) and \ref{Fig-01}(c), we consider each SSH chain is topologically trivial ($\eta = -0.5$). Subsequently, in subfigure \ref{Fig-01}(b) and \ref{Fig-01}(d), we consider individual SSH chain as nontrivial ($\eta = 0.5$). The Floquet bands are shown for cylindrical geometry, where its axis is along $x$-direction or $y$-direction. Here and throughout the paper, the axis of the cylinder along $x$-direction (or $y$-direction) means periodic boundary condition (PBC) is considered along $y$-direction (or $x$-direction), and open boundary condition (OBC) is assumed along the $x$-direction (or $y$-direction). These Floquet bands are presented in the high-frequency regime with $\Omega = 6.0$ and the driving amplitude $A_0 = 0.5$. In both cases, we obtain nontrivial topology with $C = -1$, which was topologically trivial with $C=0$ in the undriven case \cite{Agrawal_2022-02}. With the application of the CPL, the $N$ stacked SSH model is transformed into a Chern insulator. In the undriven case, we found that the topological properties remained unchanged irrespective of whether underlying SSH chains are topologically trivial or nontrivial. We observe qualitatively similar characteristics for this specific driving amplitude and frequency in both cases. To illustrate the complete behavior of the topological properties in driving parameter space, we present a phase diagram in the next subsection.

\subsection{Phase diagram}

\begin{figure}[b]
\includegraphics[width = 0.45\textwidth]{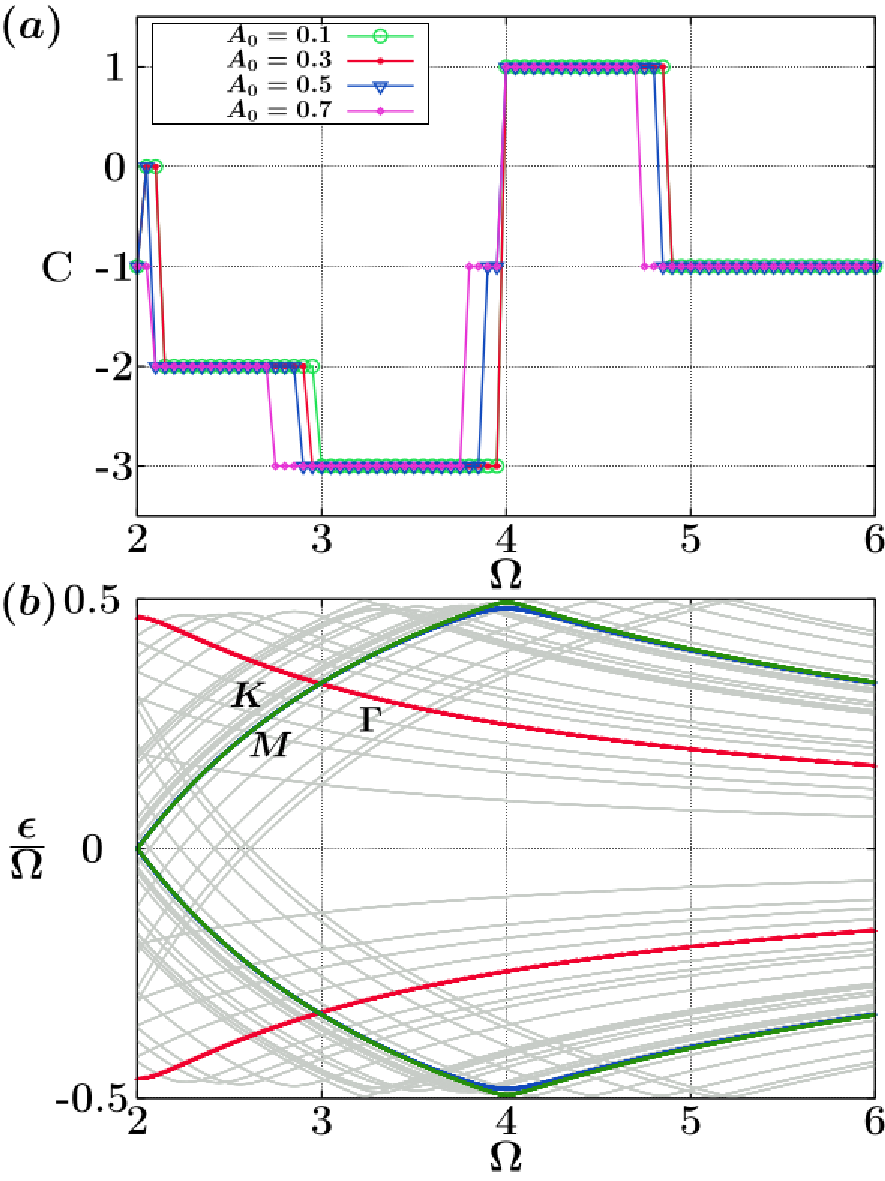}
\caption{Different topological phases with different Chern numbers $C$ are shown as the function of driving frequency $\Omega$ and driving amplitude $A_0$. In this case, we set $\eta = -0.5$; hence, the individual SSH chain is topologically trivial. Here, we consider $9$ photon sectors in the Hamiltonian to get the desired convergence. In subfigure (b), the Floquet bands are plotted in FBZ along the high symmetric path. Here, $\Gamma$ is $(\pi, \pi)$, $K$ is $(2\pi, \pi)$ and $M$ is $(2\pi, 2\pi)$.}
\label{Fig-02}
\end{figure}

\begin{figure}
\includegraphics[width = 0.45\textwidth]{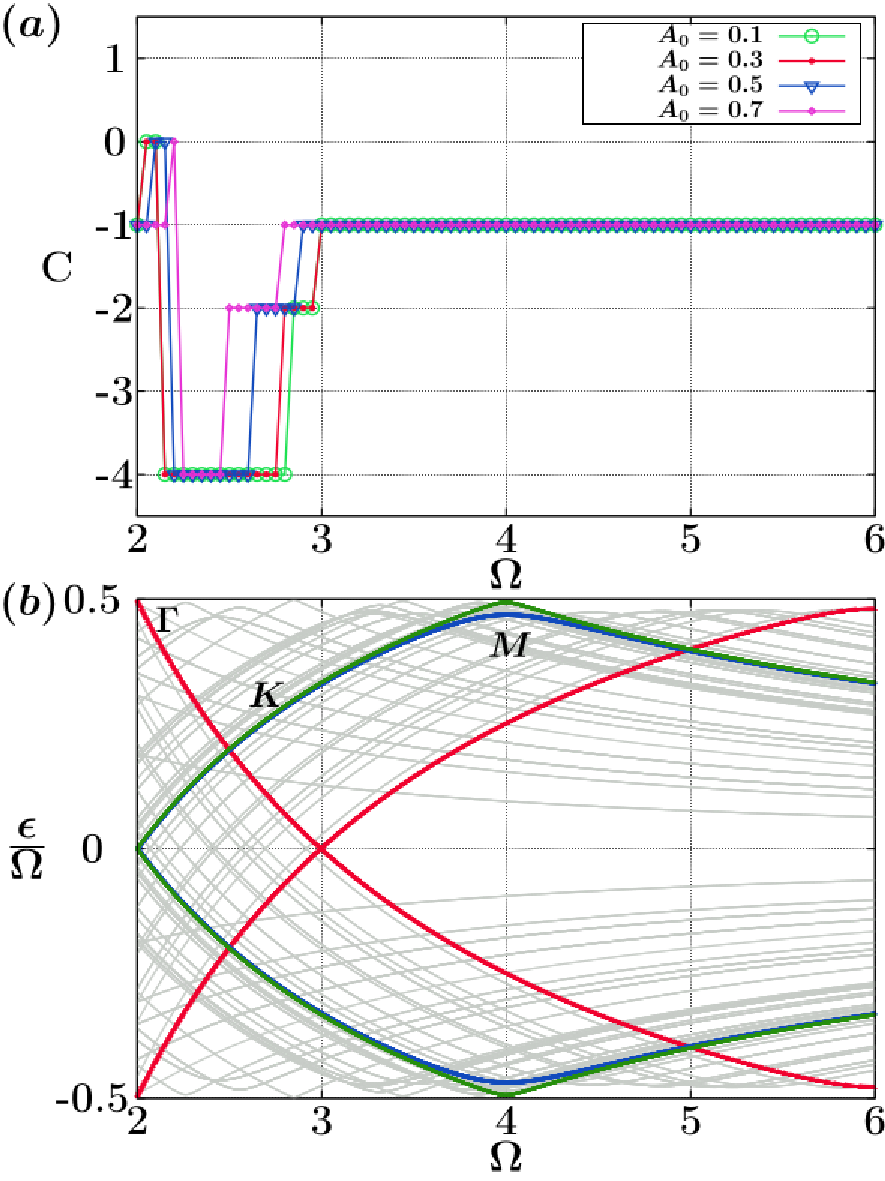}
\caption{The results presented here is similar to Fig. \ref{Fig-02}, but here we consider individual SSH chain as topologically nontrivial.}
\label{Fig-04}
\end{figure}

In this section, we demonstrate two different phase diagrams depending on the topological property of the individual SSH chain. The phase diagrams are plotted in the driving parameter space of amplitude $(A_0)$ and frequency $(\Omega)$ to illustrate various topological phases. The result for the case when the individual SSH chain is topologically trivial is presented in Fig. \ref{Fig-02}(a), whereas Fig. \ref{Fig-04}(a) shows the result when the individual SSH chains are topologically nontrivial. In these phase diagrams, we present the sum of the Chern number of all the Floquet bands below $\epsilon = 0$ as a function of the driving frequency $\Omega$ for four different values of driving amplitudes $A_0 = 0.1, 0.3, 0.5 \, \rm{and} \, 0.7$. As stated in Ref. \cite{Rudner2013}, in the frequency domain, one can calculate the number of chiral edge states in a particular gap by summing the Chern number of all the Floquet bands below that gap. 

We consider $9$ photon sectors for all the calculations to achieve the convergence. When the individual SSH chain is topologically trivial, we observe five topological phases with $C = 0,\pm1,-2,$ and $-3$. In the low-frequency regime ($\Omega < 4.0$), we obtain topological phases with the high Chern number, while in the high-frequency regime ($\Omega > 4.0$), the system exhibits topological phases with $C = \pm 1$. In Fig. \ref{Fig-04}, where the phase diagram is plotted for the nontrivial case, we observe four different topological phases with Chern number $C = 0, -1, -2,$ and $-4$. For this case, the system saturates at the topological phase with $C = -1$ in the high-frequency regime. It is important to note that, for the different values of the driving amplitude $A_0 \in [0.1, 0.7]$, the variation in the Chern number with the driving frequency $\Omega$ is almost similar. The value of $\Omega$ determines the variation in the Chern number. Nevertheless, both the phase diagrams show that, compared to the undriven case, the Floquet version of the $N$ stacked SSH chains displays much richer topological phases with high Chern numbers. 

It is well known that the bulk boundary correspondence in the Floquet system is not the same as the undriven cases. In the case of the undriven systems, the edge states appear only in the energy gaps between the bulk bands. However, in the Floquet version of these systems, we have infinite copies of the undriven systems. The driving affects not only the band gap between the original bands but also different copies or replicas. Consequently, edge states can also appear in between Floquet replica bands. Here, we consider the Floquet bands or quasi-energy bands only in the first FBZ, where the edge states can exist at the central gap $\frac{\epsilon}{\Omega} = 0$; and as well as at the boundary of the FBZ, i.e., at $\frac{\epsilon}{\Omega} = 0.5$. As a consequence, the total number of chiral edge states is calculated from the relation $C = C_0 - C_{\pi}$, where $C_0$ and $C_{\pi}$ respectively measure the number of chiral edge states in the central gap and the gap around the Floquet zone boundary. For the frequency regime $2.0 \leq \Omega \leq 6.0$, we observe $C_{\pi} = 0$ at the Floquet zone boundary. Therefore, the Chern number is always $C = C_0$ in this frequency regime.

The topological phase transitions observed in the phase diagrams occur in the system because of the closing and re-opening of some of the band gaps. Therefore, in Figs. \ref{Fig-02}(b) and \ref{Fig-04}(b), we show the bands in the first FBZ along the high symmetric path. Earlier, we mentioned that the topological phases are almost independent of $A_0$; hence, we plot these band diagrams only for $A_0 = 0.1$. The phase transition occurs due to the band gap closing at $\epsilon/\Omega=0$ or at the Floquet zone boundary $\epsilon/\Omega= \pm 0.5$. Since the maximum band gap in the undriven system is $4.0$, the band gap closing occurs at $\epsilon/\Omega= \pm 0.5$ when $\Omega = 4.0$. 

\subsection{Demonstration of the edge states of the Floquet topological phases with high Chern number}

\begin{figure}[t]
\includegraphics[width = 0.45\textwidth]{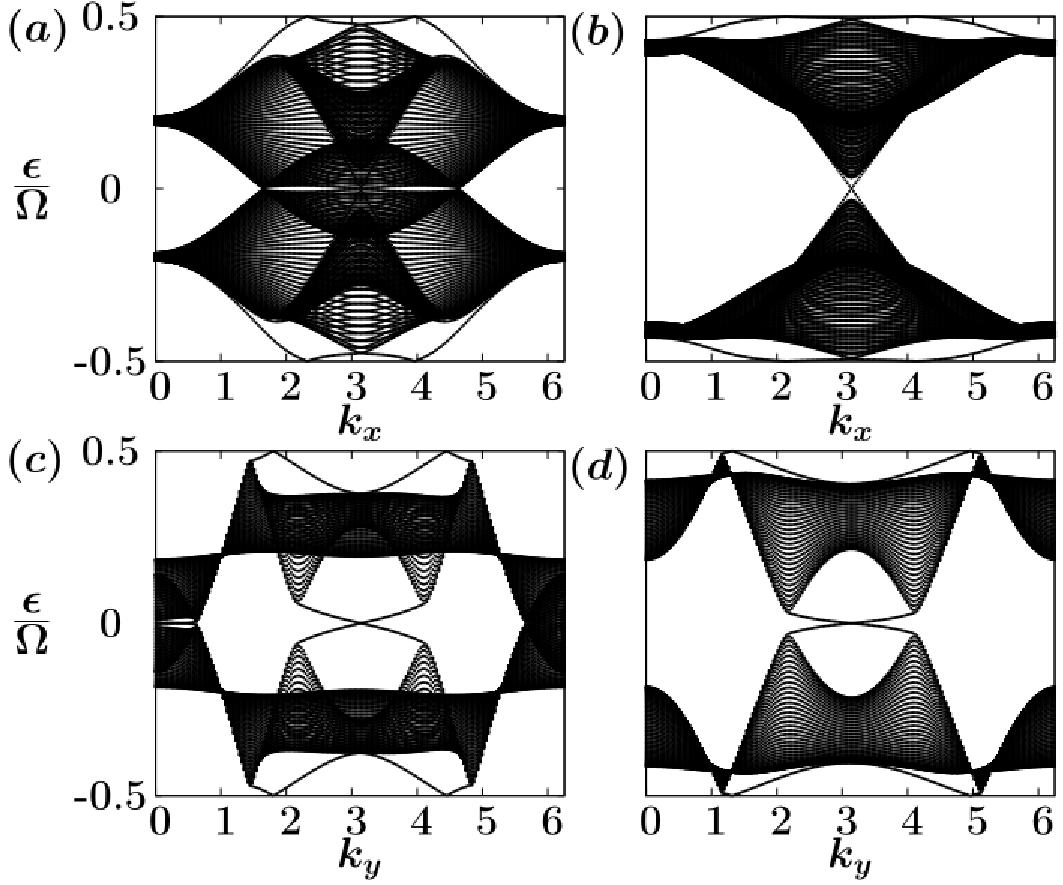}
\caption{The quasienergy bands for cylindrical geometry are shown for the Chern numbers $C = -2$ and $C = -3$ corresponding to the phase diagram Fig. \ref{Fig-02}. In subfigures (a) and (c), the bands are plotted for the driving amplitude $A_0 = 0.5$ and the driving frequency $\Omega = 2.5$. In subfigures (b) and (d), the bands are plotted for $A_0 = 0.5$ and $\Omega = 3.5$. In both cases, the individual SSH chain is considered topologically trivial.}
\label{Fig-03}
\end{figure}

\begin{figure}[b]
\includegraphics[width = 0.45\textwidth]{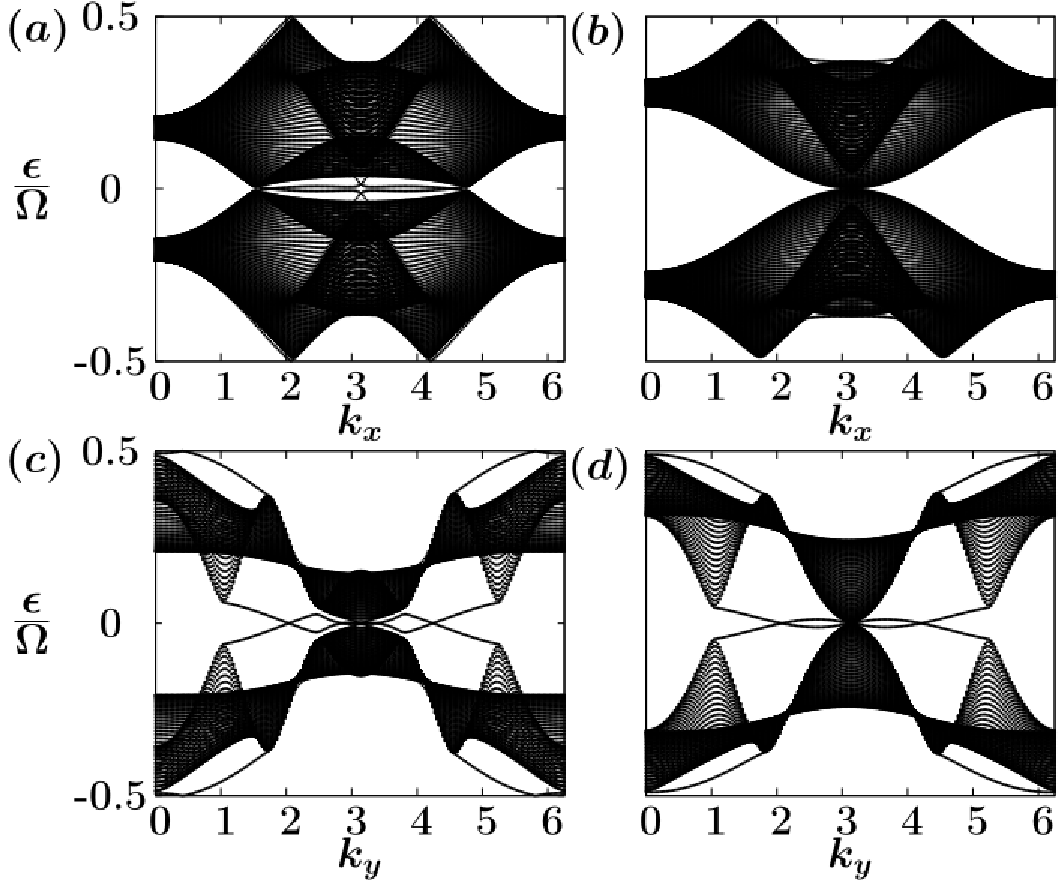}
\caption{The quasienergy bands for cylindrical geometry are shown for the Chern numbers $C = -4$ and $C = -2$ corresponding to the phase diagram presented in Fig. \ref{Fig-04}. In subfigures (a) and (c), the bands are shown for the driving amplitude $A_0 = 0.5$ and the driving frequency $\Omega = 2.5$. In subfigures (b) and (d), the bands are presented for $A_0 = 0.5$ and $\Omega = 2.9$. In both cases, the individual SSH chain is considered as topologically nontrivial.}
\label{Fig-05}
\end{figure}

This section discusses the edge states observed in the energy band diagrams for the higher Chern numbers with $|C| > 1$. In Fig. \ref{Fig-02}, we observed topological phases with the Chern number $C = -2$ and $-3$. Here, we select two pairs of the driving amplitudes and frequencies $(A_0 = 0.5, \Omega = 2.5)$ and $(A_0 = 0.5, \Omega = 3.5$) from the phase diagram, where the corresponding Chern numbers are $C = -2$ and $-3$, respectively. We choose $A_0 = 0.5$ so that the band gap is prominent and the edge states can be clearly visible. In Fig. \ref{Fig-03}(a)-(d), we have shown the energy band diagram for the case of cylindrical geometry. In this figure, we consider individual SSH chains to be topologically trivial. In Fig. \ref{Fig-05}(a)-(d), the energy band diagrams for cylindrical geometry are presented, where the individual SSH chain is nontrivial. These figures show the edge states for the topological phases with $C = -4$ and $C = -2$. The band diagrams (Figs. \ref{Fig-03} and \ref{Fig-05}) exhibit edge states along both directions; thus, these show an actual 2D-like system even though the system is constructed as a weak topological insulator by stacking many SSH chains.

\section{\label{Sec:Low-energy-H}Low-energy Hamiltonian}

\begin{figure}
\includegraphics[width = 0.5\textwidth]{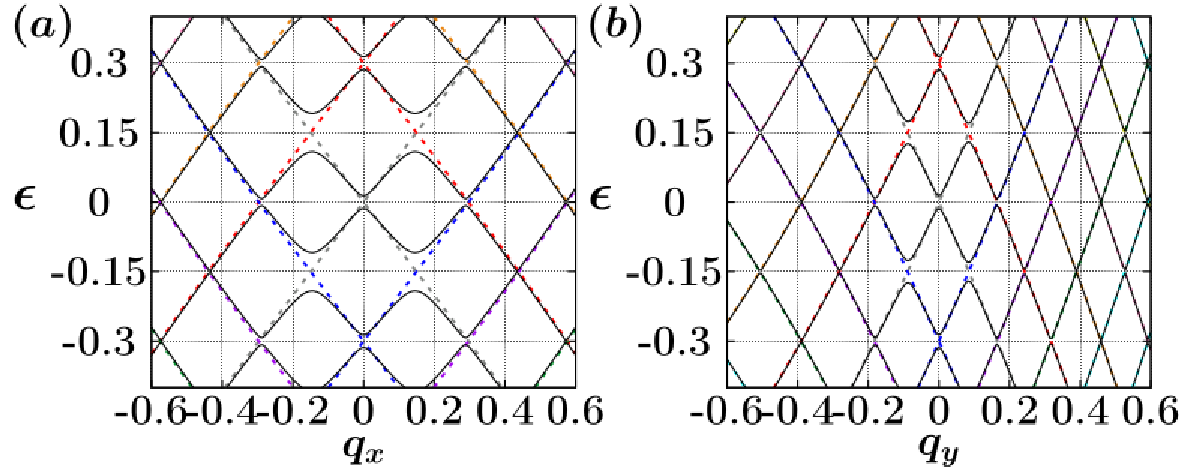}
\caption{The quasienergy bands repulsion in the case of the low-energy Hamiltonian is shown for the cylindrical geometry. In subfigure (a), the cylinder's axis is considered along the $y$-direction. The boundary condition is interchanged in subfigure (b), i.e., the PBC is considered along the $x$-direction, and the OBC is considered along the $y$-direction. Here, the individual SSH chain is considered as nontrivial. The colored dashed lines show the quasienergy bands for the undriven system, whereas the black solid lines are used for the driven case. We set the driving parameters at $\Omega = 0.3$ and $A_0 = 0.05$.}
\label{Fig-low-energy-NTri}
\end{figure}

\begin{figure}[b]
\includegraphics[width = 0.5\textwidth]{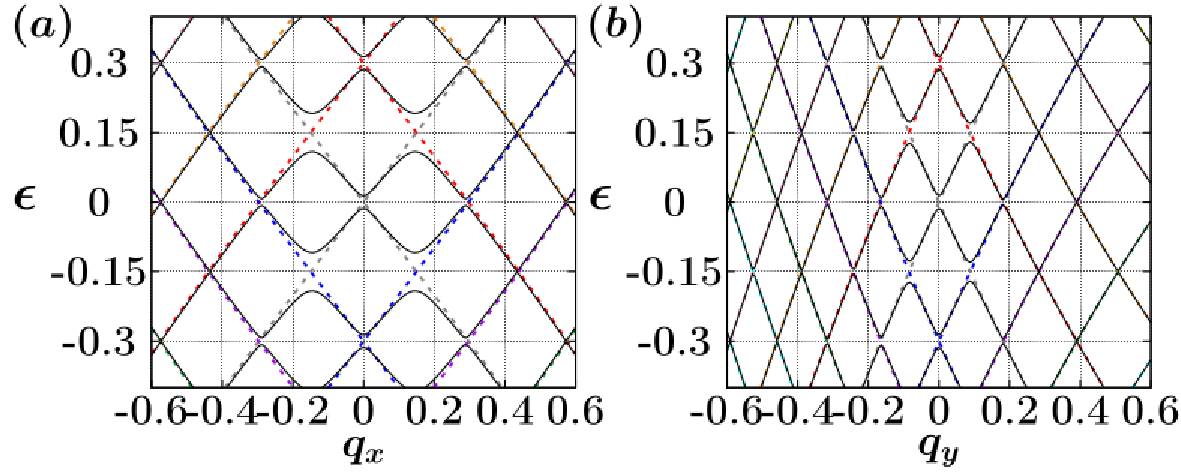}
\caption{The results presented here are similar to Fig. \ref{Fig-low-energy-NTri}, but here the individual SSH chain is topologically trivial.}
\label{Fig-low-energy-Tri}
\end{figure}

The low-energy Hamiltonian near the band touching point $\bs{K}_D = \left[ \pi, \cos^{-1}\frac{\eta}{\delta}\right]$, can be derived by substituting $\bs{k} = \bs{q} + \bs{K}_D $ in Eq. (\ref{Eq-NSSH}). Here, $|\bs{q}| \ll 1$ and under this condition the Hamiltonian will be
\begin{equation}
\begin{split}
h_x^{(0)} &= \frac{q_x^2}{2} - \eta q_y^2 - 2 q_y \sqrt{\delta^2 - \eta^2}\\
h_y^{(0)} & = -q_x - q_x q_y \sqrt{\delta^2 - \eta^2}
\end{split}
\label{Low-E-H-CPL}
\end{equation}
For the driven case, we obtain Fourier components of the low energy Hamiltonian by substituting $\bs{q} \rightarrow \bs{q} + \bs{A}(t)$
\begin{equation}
\begin{split}
h_x^{(1)} &= q_x \frac{A_0}{2} + i \eta q_y A_0 + i A_0 \sqrt{\delta^2 - \eta^2}\\
h_y^{(1)} & = -\frac{A_0}{2} + i q_x \frac{A_0}{2} + q_y \frac{A_0}{2} \sqrt{\delta^2 - \eta^2} \\
h_x^{(2)} & = \frac{A_0^2}{8} + \eta \frac{A_0^2}{4}\\
h_y^{(2)} & = i \frac{A_0^2}{4} \sqrt{\delta^2 - \eta^2}
\end{split}
\label{Low-E-drivenH}
\end{equation}
Here, we see that, unlike graphene, the dispersion relation of the low-energy Hamiltonian of the $N$-stacked SSH model is still unconventional, having quadratic dispersion with asymmetry along both $k_x$ and $k_y$ directions due to the presence of a coupling term. The unconventional dispersion makes the system more complex than any other 2D system. We now investigate in detail the behavior of the low-energy Hamiltonian.

The effect of the asymmetry in the Hamiltonian can be seen in Figs. \ref{Fig-low-energy-NTri} and \ref{Fig-low-energy-Tri}, where we have projected the Floquet bands along one of the directions of the quasi momenta and set the value of the other quasi momentum equals zero. In this figure, the dotted lines represent energy bands of the undriven system, where different colors denote different photon sectors. The black solid lines are used for the driven case. As we turn on the driving, the Floquet bands corresponding to different photon sectors repel each other and create band gaps where Floquet edge states can appear. Our primary goal is to investigate whether the Floquet bands of the $N$ stacked model have the same hierarchical structure as observed in graphene \cite{FoaTorres2015}. Even though our system is very different from graphene, we observe some hierarchical structure in the Floquet band gaps at $\epsilon=0$, and also at $\epsilon = \pm \frac{\Omega}{2}$. We observe that the behavior of the level repulsion around the central gap and the Floquet zone boundary is qualitatively similar to graphene. Here, the width of the Floquet gaps is approximately of the order of $\left(\frac{A_0}{\Omega}\right)^{\Delta m}$, where $\Delta m$ is the difference between the photon sectors. 

In Figs. \ref{Fig-low-energy-NTri} and \ref{Fig-low-energy-Tri}, we show by red and blue dotted lines that the largest Floquet band gap occurs at $\epsilon = \pm \frac{\Omega}{2}$, due to the repulsion between the bands with photon sectors $m = \pm 1$ and $m = 0$. However, the width of the Floquet band gap at $\epsilon = 0$ between the bands in the zero photon sector is of the order of $\left(\frac{A_0}{\Omega}\right)^2$ \cite{FoaTorres2015}.

\section{\label{Sec:LPL} ${\bs N}$ stacked SSH chains model under linearly polarized light}

\begin{figure}
\includegraphics[width = 0.45\textwidth]{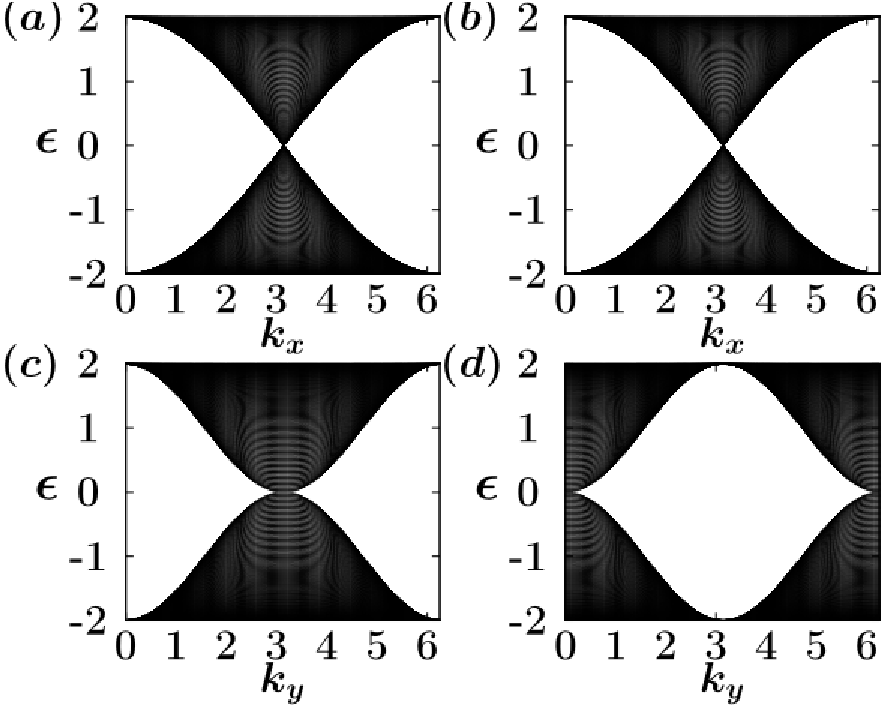}
\caption{The quasienergy bands with a semi-Dirac point are shown for the undriven case with $\delta = |\eta|$. In subfigures (a) and (c), we set $\delta = 0.5$ and $\eta = -0.5$, and hence the individual SSH chain is topologically trivial. On the other hand, in subfigures (b) and (d), the individual SSH chain is made topologically nontrivial by setting $\delta = 0.5$ and $\eta = 0.5$. }
\label{Fig-undriven-Semi-Dirac}
\end{figure}

Earlier, we have shown that the $N$ stacked SSH model has non-linear dispersion. Therefore, we choose the system parameters so that this model also shows a semi-Dirac point, i.e., a band touching point where the dispersion along one direction is linear and quadratic along the other. A similar semi-Dirac point is observed in the hexagonal lattice. The semi-Dirac point of this system is split into two Dirac points if the system is shined by an LPL \cite{CPL-06}. For our study,  
we consider two cases of the LPL: along $x$ and $y$-directions. Our goal is to observe whether the semi-Dirac point of the $N$ stacked SSH model also splits into two Dirac points. Here, we set $\delta = |\eta|$ in Eq. (\ref{Eq-NSSH}). When $N$ stacked SSH chain is constructed with trivial SSH chains, we observe a semi-Dirac point at $\left[\pi, \pi\right]$ as shown in Figs. \ref{Fig-undriven-Semi-Dirac}(a) and \ref{Fig-undriven-Semi-Dirac}(c). For the other case, when $N$ stacked SSH chain is constructed with nontrivial SSH chains, we find the semi-Dirac point at $\left[\pi, 0\right]$ (or at $\left[\pi, 2\pi\right]$) as shown in Figs. \ref{Fig-undriven-Semi-Dirac}(b) and \ref{Fig-undriven-Semi-Dirac}(d). The emergence of the semi-Dirac behavior can be identified by deriving the low-energy Hamiltonian around these band touching points as follows:
\begin{equation}
h_x^{(0)} = \frac{q_x^2}{2} - \eta q_y^2;~ 
h_y^{(0)} = -q_x
\label{Low-E-H-LPL}
\end{equation}
The above low energy Hamiltonian of the $N$ stacked SSH model shows semi-Dirac-like behavior for the condition $\left[ \delta = |\eta| \right]$. We now separately study the role of LPL along $x$ and $y$-directions. 

\begin{figure}
\includegraphics[width = 0.45\textwidth]{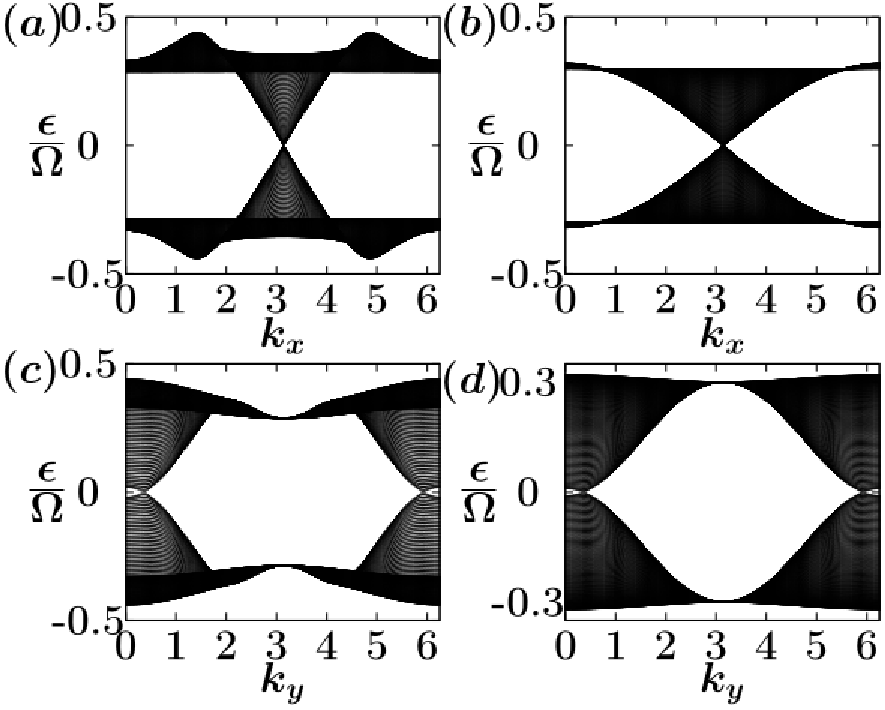}
\caption{The quasienergy bands are presented when the LPL is applied along the $x$-direction for the case when the individual SSH chain is topologically nontrivial. We fix the driving parameters at $A_0 = 0.5$ and $\delta = \eta = 0.5$ in all the subfigures. In subfigures (a) and (c), we set $\Omega = 3.0$, while in subfigures (b) and (d), we set $\Omega = 6.0$. Here, we observe the emergence of the semi-Dirac-like point as LPL is applied along $x$-direction.}
\label{Fig-LPL-x-NTri-01}
\end{figure}
\begin{figure}[t]
\includegraphics[width = 0.45\textwidth]{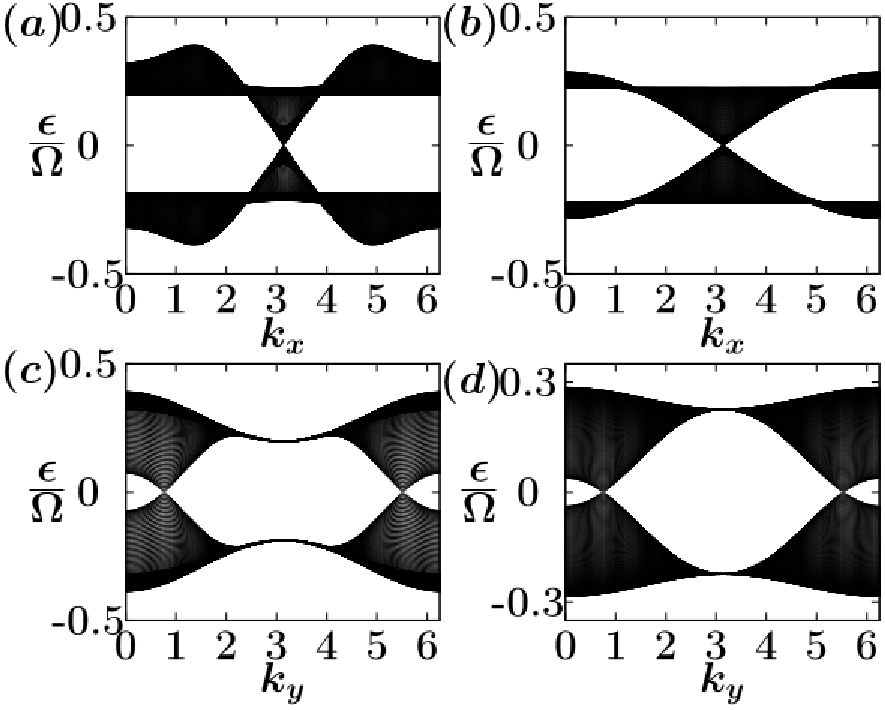}
\caption{This figure is similar to Fig. \ref{Fig-LPL-x-NTri-01}, but here we consider stronger driving amplitude $A_0 = 1.0$.}
\label{Fig-LPL-x-NTri-02}
\end{figure}

\subsection{Linearly polarized light along ${\bs x}$-direction}

\begin{figure}
\includegraphics[width = 0.45\textwidth]{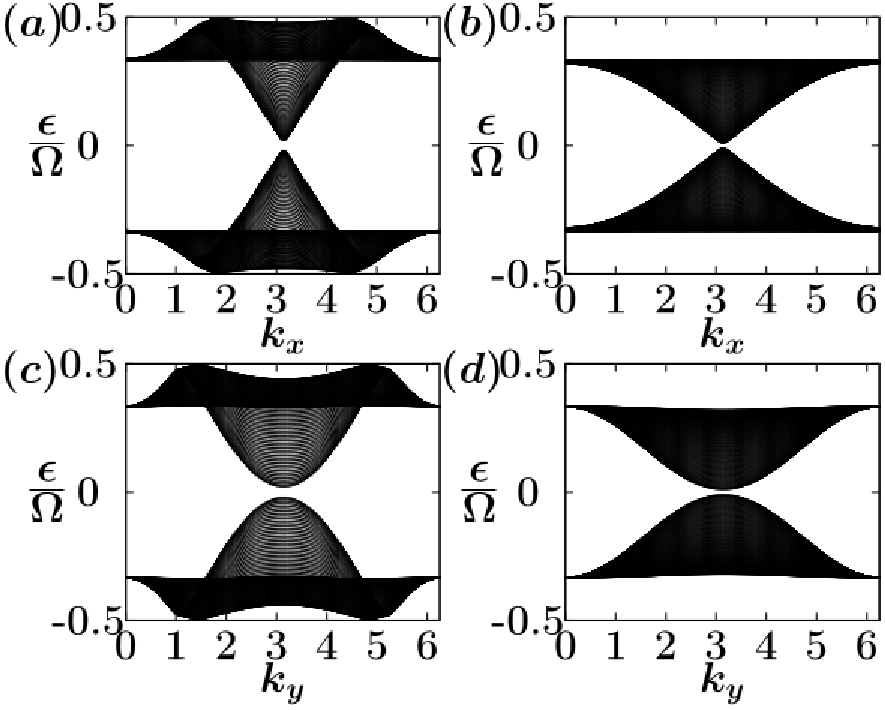}
\caption{The quasienergy bands are presented for the case when the LPL is applied along the $x$-direction. Here, we fix the parameters $\delta = -\eta = 0.5$; hence, the individual SSH chain is topologically trivial. We set the driving amplitude $A_0 = 0.5$. In subfigures (a) and (c), we set $\Omega = 3.0$, while in subfigures (b) and (d), we set $\Omega = 6.0$. Here, we see a gap opening in the quasienergy band at the semi-Dirac-like point.}
\label{Fig-LPL-x-Tri-01}
\end{figure}

First, we consider the case when the LPL is applied along $x$-direction of the form $A(t) = A_0 \cos \Omega t$. For this driving, we find the Fourier component of the Hamiltonian for $m=0$ as:
\begin{subequations}
\begin{equation}
H^{(0)}_{N-SSH} = \, \bs{d}^{(0)}_{N-SSH} \cdot \bs{\sigma}
\end{equation}
where
\begin{equation}
\begin{split}
\left(d^{(0)}_{N-SSH}\right)_x = & \, (1-\eta) + (1+\eta) \, \cos k_x \, J_0(A_0) + \delta \, \cos k_y, \\
&- \delta \, \cos k_x \, \cos k_y \, J_0(A_0)
\end{split}
\end{equation}
\begin{equation}
 \left(d^{(0)}_{N-SSH}\right)_y = \, (1+\eta) \, \sin k_x \, J_0(A_0) - \delta \, \sin k_x \, \cos k_y \, J_0(A_0).
 \end{equation}
\end{subequations}
\noindent The Fourier component of the Hamiltonian for $m=1$ is obtained as:
\begin{subequations}
\begin{equation}
H^{(1)}_{N-SSH} = \, \bs{d}^{(1)}_{N-SSH} \cdot \bs{\sigma}
\end{equation}
where
\begin{equation}
\left(d^{(1)}_{N-SSH}\right)_x = \, -(1+\eta) \, \sin k_x \, J_1(A_0) + \delta \sin k_x \cos k_y J_1(A_0) 
\end{equation}
\begin{equation}
\left(d^{(1)}_{N-SSH}\right)_y = \, \left[(1+\eta) - \delta \cos k_y \right] \, \cos k_x \, J_1(A_0) 
\end{equation}
\end{subequations}
\noindent and for $m=2$ as:
\begin{subequations}
\begin{equation}
H^{(2)}_{N-SSH} = \, \bs{d}^{(2)}_{N-SSH} \cdot \bs{\sigma}
\end{equation}
\begin{equation}
\left(d^{(2)}_{N-SSH}\right)_x = \, -(1+\eta) \, \cos k_x \, J_2(A_0) + \delta \cos k_x \cos k_y J_2(A_0)
\end{equation}
\begin{equation}
 \left(d^{(2)}_{N-SSH}\right)_y = \, - \left[ (1+\eta) - \delta \cos k_y \right] \, \sin k_x J_2(A_0).
\end{equation}
\end{subequations}
The Floquet energy band diagrams are shown in Figs. \ref{Fig-LPL-x-NTri-01}, \ref{Fig-LPL-x-NTri-02} and \ref{Fig-LPL-x-Tri-01}. Here again, we consider two different cases depending on the topological property of the individual SSH chain. In Figs. \ref{Fig-LPL-x-NTri-01} and \ref{Fig-LPL-x-NTri-02}, each SSH chain is considered as nontrivial by setting $\eta = 0.5$. Unlike the case of a hexagonal lattice, here we observe that the semi-Dirac-like point splits into two band-touching points with non-linear dispersion along $y$-direction. This behavior is observed for both high and low-frequency regimes. In Figs. \ref{Fig-LPL-x-NTri-01}(a) and (c), we set the driving frequency $\Omega = 3.0$. On the other hand, in Figs. \ref{Fig-LPL-x-NTri-01}(b) and (d), we set $\Omega = 6.0$. The driving amplitude is fixed at $A_0 = 0.5$ for both frequencies. As we increase the driving amplitude, the separation between the two band touching points increases, as shown in Fig. \ref{Fig-LPL-x-NTri-02}. On the contrary, when we consider each SSH chain as topologically trivial, the semi-Dirac-like point does not split, but a band gap opens at that point. This result is shown in Fig. \ref{Fig-LPL-x-Tri-01}. 

\subsection{Linearly polarized light along ${\bs y}$-direction}

\begin{figure}[b]
\includegraphics[width = 0.45\textwidth]{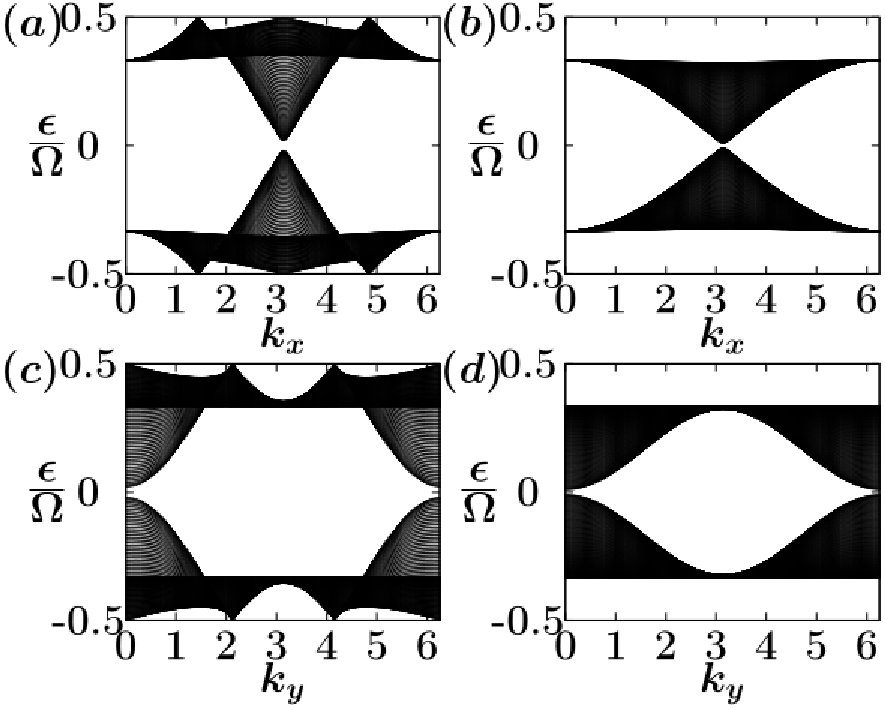}
\caption{The quasi-energy bands are shown for the case when the LPL is applied along $y$-direction, and here we consider the individual SSH chain as topologically nontrivial by setting the parameters $\delta = \eta = 0.5$. The driving amplitude is set at $A_0 = 0.5$ in all the subfigures. In the subfigures (a) and (c), we consider $\Omega = 3.0$, while $\Omega = 6.0$ is considered in subfigures (b) and (d). Here, we also observe a band gap opening at the semi-Dirac-like point.}
\label{Fig-LPL-y-NTri-01}
\end{figure}

\begin{figure}[t]
\includegraphics[width = 0.45\textwidth]{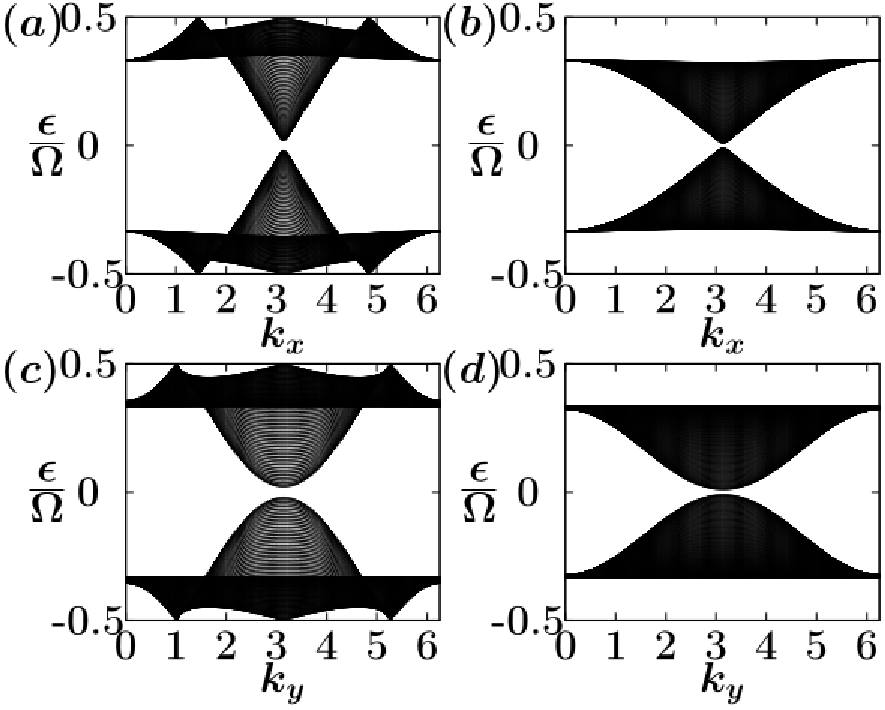}
\caption{Similar results as Fig. \ref{Fig-LPL-y-NTri-01} are presented, but here, the individual SSH chain is considered topologically trivial by setting $\delta = -\eta = 0.5$.}
\label{Fig-LPL-y-Tri-01}
\end{figure}

We now consider the other case where the LPL is applied along $y$-direction of the same form as earlier, i.e., $A(t) = A_0 \cos \Omega t$. Like the previous case, for this driving, we find the Fourier component of the Hamiltonian for $m=0$ as:
\begin{subequations}
\begin{equation}
H^{(0)}_{N-SSH} = \, \bs{d}^{(0)}_{N-SSH} \cdot \bs{\sigma}
\end{equation}
where
\begin{equation}
\begin{split}
\left(d^{(0)}_{N-SSH}\right)_x &= \, (1-\eta) + (1+\eta) \, \cos k_x + \delta \, \cos k_y \, J_0(A_0)\\
&- \delta \, \cos k_x \, \cos k_y \, J_0(A_0)
\end{split}
\end{equation}
\begin{equation}
 \left(d^{(0)}_{N-SSH}\right)_y = \, (1+\eta) \, \sin k_x - \delta \, \sin k_x \, \cos k_y \, J_0(A_0).
\end{equation}
\end{subequations}
\noindent Again, we derive the Fourier component of the Hamiltonian for $m=1$ as:
\begin{subequations}
\begin{equation}
H^{(1)}_{N-SSH} = \, \bs{d}^{(1)}_{N-SSH} \cdot \bs{\sigma}
\end{equation}
\begin{equation}
\left(d^{(1)}_{N-SSH}\right)_x = \, -\delta \, \sin k_y \, J_1(A_0) + \delta \cos k_x \sin k_y J_1(A_0) 
\end{equation}
\begin{equation}
\left(d^{(1)}_{N-SSH}\right)_y = \, \delta \sin k_x \, \sin k_y \, J_1(A_0) 
\end{equation}
\end{subequations}
\noindent and for $m=2$ as:
\begin{subequations}
\begin{equation}
H^{(2)}_{N-SSH} = \, \bs{d}^{(2)}_{N-SSH} \cdot \bs{\sigma}
\end{equation}
\begin{equation}
\left(d^{(2)}_{N-SSH}\right)_x = \, -\delta \, \cos k_y \, J_2(A_0) + \delta \cos k_x \cos k_y J_2(A_0)  
\end{equation}
\begin{equation}
 \left(d^{(2)}_{N-SSH}\right)_y = \, \delta \sin k_x \cos k_y \, J_2(A_0).
\end{equation}
\end{subequations}
In Fig. \ref{Fig-LPL-y-NTri-01}, the Floquet band diagrams are shown for the case where the individual SSH chain is topologically nontrivial. On the other hand, in Fig. \ref{Fig-LPL-y-Tri-01}, the Floquet band diagrams are presented for the trivial case. For both cases, the semi-Dirac-like point does not split, and a band gap opens at that point.

\section{\label{Sec:Discussion}Summary}

We study the effect of time-periodic driving on the $N$ stacked SSH model and examine the topological properties of this system. We use circularly polarized light as a periodic drive and compute the effective Hamiltonian using the Floquet replica method. This method is exact, and one can determine the topological phases in high-frequency and low-frequency regimes. This system exhibits topological phases with high Chern numbers in the low-frequency regime. We consider two different versions of the $N$ stacked SSH model: in one case, the individual SSH chain is topologically trivial, and the chains are nontrivial in the other case. Under periodic driving, both versions have distinct topological properties. From the phase diagrams in the driving parameter space, we observe topological phases with high ($|C| > 1$) Chern numbers in the low-frequency regime, whereas $|C|=1$ is observed for the high-frequency cases. When the $N$ stacked SSH model has all trivial SSH chains, the highest Chern number is $C = -3$, while for each nontrivial SSH chain, the highest Chern number is $C = -4$. We have also observed that, with the varying driving amplitude $A_0$, the topological phases remain the same but change with the varying driving frequency $\Omega$. The topological transition occurs when a band gap closing or reopening happens in the system. Therefore, with the varying $\Omega$, we have shown that the quasienergy bands lie in the first FBZ. We have also shown the band diagrams in cylindrical geometry to present the results with the high Chern number. 

In the Floquet systems, the total number of chiral edge states between the gap at $\epsilon=0$ and $\epsilon = \frac{\Omega}{2}$ is calculated from the relation $C = C_0 - C_{\pi}$. Here, $C_0$ and $C_{\pi}$ are the sum of the Chern number of all the Floquet bands below the energy at $\epsilon=0$ and $\epsilon = \frac{\Omega}{2}$, respectively. For both cases, we always find $C_{\pi} = 0$, which gives $C = C_0$. The model we study in this paper is nontrivial due to the presence of coupling between the quasimomenta $k_x$ and $k_y$. We construct the low-energy Hamiltonian around the band touching points to observe the interplay of this coupling and the periodic driving more prominently. The $k_x-k_y$ coupling term makes the dispersion relation of this system unconventional. Even after the presence of a very different dispersion relation than graphene, the $N$ stacked SSH model shares a similar signature of hierarchy in the Floquet gaps with graphene.

Due to the quadratic dispersion, we observe a semi-Dirac-like band touching point for a particular choice of the system parameter. We find that, for the periodic driving with LPL along $x$-direction, the semi-Dirac-like point splits into two band touching points with nonlinear dispersion. On the other hand, when the LPL is applied along the $y$-direction, a band gap opens around the semi-Dirac-like point. The same observation is shown for the hexagonal lattice \cite{CPL-06}. We have shown the band diagrams for the existence of the semi-Dirac point, splitting into two band points and the opening of the band gap for the cylindrical geometry.

\begin{acknowledgments}
Authors acknowledge financial support from DST-SERB, India, through the Core Research Grant CRG/2020/001701 and also through MATRICS Grant No. MTR/2022/000691.
\end{acknowledgments}

\bibliography{References}

\end{document}